\documentclass[aps,prl,twocolumn,superscriptaddress]{revtex4-2}

\usepackage{amsmath,amssymb,braket,bm}
\usepackage{bbm,times,mathtools,xcolor}
\usepackage[colorlinks=true,allcolors=blue]{hyperref}

\newcommand{\Tr}{\mathrm{Tr}}

\begin{document}

\title{The Thermodynamic Geometry of Conditional Control}

\author{J. G. G. de Oliveira, Jr.\href{https://orcid.org/0000-0002-1490-8859}{\includegraphics[scale=0.05]{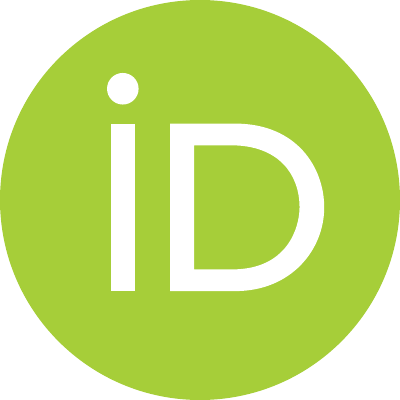}}}
\affiliation{NEMeS, Department of Exact Sciences, State University of Santa Cruz, Ilhéus, Bahia, 45662-900, Brazil}
\affiliation{Department of Physics, Federal University of Paran\'a, Curitiba, Paran\'a, P.O. Box 19044, 81531-980, Brazil}

\author{M. E. R. Filippetto\href{https://orcid.org/0009-0002-9682-7880}{\includegraphics[scale=0.05]{orcidid.pdf}}}
\affiliation{Department of Physics, Federal University of Paran\'a, Curitiba, Paran\'a, P.O. Box 19044, 81531-980, Brazil}
\affiliation{Department of Physics \& Astronomy, University of Manchester, Manchester, UK}

\author{T. A. B. Pinto Silva
\href{https://orcid.org/0009-0003-3677-8842}{\includegraphics[scale=0.05]{orcidid.pdf}}}
\affiliation{Schulich Faculty of Chemistry and Helen Diller Quantum Center, Technion-Israel Institute of Technology, Haifa 3200003, Israel}

\author{A. C. S. Costa\href{https://orcid.org/0000-0002-4014-0695}{\includegraphics[scale=0.05]{orcidid.pdf}}}
\affiliation{Department of Physics, Federal University of Paran\'a, Curitiba, Paran\'a, P.O. Box 19044, 81531-980, Brazil}

\author{R. M. Angelo\href{https://orcid.org/0000-0002-7832-9821}{\includegraphics[scale=0.05]{orcidid.pdf}}}
\affiliation{Department of Physics, Federal University of Paran\'a, Curitiba, Paran\'a, P.O. Box 19044, 81531-980, Brazil}

\date{\today}

\begin{abstract}
Information is widely regarded as the resource underlying the thermodynamic advantages enabled by conditional control. We show, however, that informational quantities such as Holevo information and accessible distinguishability, although constraining the achievable advantage, do not uniquely determine its thermodynamic value. The missing ingredient is a passive spectral rearrangement vector that characterizes the effect of conditioning on the ensemble. Specifically, the conditional-control advantage is exactly determined by the geometric pairing between this vector and the Hamiltonian energy-gap structure. This result reveals a thermodynamic geometry of conditional control, explains how informationally equivalent ensembles can possess different thermodynamic values, and identifies the passive spectral rearrangement vector as the minimal operational descriptor required to determine the thermodynamic value of conditional control for a fixed Hamiltonian.
\end{abstract}

\maketitle

\textit{Introduction.}---The intimate connection between information and thermodynamics has shaped our understanding of physical processes for more than six decades. Since Landauer established the physical cost of information processing and Bennett clarified its role in reversible computation, information has come to be regarded as a genuine physical resource capable of constraining energetic transformations and work extraction processes~\cite{Landauer,Bennett}. This perspective has motivated extensive research on Maxwell demons, feedback-controlled systems, and information-assisted thermodynamics, both in classical and quantum regimes~\cite{Goold2016,VinjanampathyAnders2016,Campaioli2024}.

In quantum thermodynamics, information is often associated with the ability to improve energetic performance through measurement and feedback~\cite{SagawaUeda2010,SagawaUeda2012}. This viewpoint underlies a broad family of quantum Maxwell demons, daemonic protocols, and information-assisted thermal machines, where information acquired about a system is subsequently converted into useful work through conditional operations~\cite{Alhambra2016,Francica2017,Elouard2017,Oftelie2026}. The general intuition is that the better one can resolve the relevant conditional branches, the more useful the available control becomes. An important question, however, is whether informational quantities are sufficient to determine the thermodynamic value of such conditional control.

At the same time, the thermodynamics of finite quantum systems has revealed the fundamental role played by passivity and ergotropy~\cite{PuszWoronowicz1978,Lenard1978,Allahverdyan2004}. These concepts provide a precise characterization of the work extractable through cyclic unitary operations and have become central tools in modern quantum thermodynamics. They also provide a natural framework for quantifying the energetic value of conditional-control protocols.

A natural question is whether the thermodynamic advantages commonly associated with Maxwell demons can be achieved without measurement readout, postselection, or classical feedback. The framework considered here addresses this question through a fully coherent protocol in which information is generated and exploited via unitary interactions alone. In this setting, the resulting work gain naturally separates into an average-state contribution and a contribution associated with branch-resolved control. The latter governs the genuinely conditional component of the weak-ergotropic gain and therefore provides a natural arena for investigating the thermodynamic value of conditional control.

Because conditional control relies on the distinguishability of conditional branches, it is natural to expect information-theoretic quantities to determine its thermodynamic value. Indeed, measures such as accessible distinguishability and Holevo information quantify the information available for conditional decision making and provide important constraints on the energetic advantage enabled by conditional operations. This viewpoint underlies much of the modern understanding of information-assisted work extraction and Maxwell-demon protocols~\cite{SagawaUeda2010,SagawaUeda2012,Alhambra2016,Francica2017}.

More recently, developments in quantum thermodynamics have revealed the central role of spectral ordering and majorization in characterizing state transformations and energetic resources~\cite{HorodeckiOppenheim2013,Brandao2015SecondLaws,Gour2018QuantumMajorization,MarshallOlkinArnold2011}. These advances suggest that thermodynamic value may depend not only on how much information is available, but also on how that information is spectrally organized within the conditional ensemble.

In this Letter we show that information alone does not determine the thermodynamic value of conditional control. While informational quantities constrain the advantage enabled by conditional control, they do not uniquely determine it. We identify the missing structure and show that the thermodynamic value of conditional control is governed by a majorization-based thermodynamic geometry that emerges naturally from the passive rearrangements induced by conditioning. This result reveals a previously unnoticed geometric structure underlying conditional control and establishes a clear distinction between informational content and thermodynamic value.

\textit{Coherent Demon Protocol and Conditional Control}.---We consider a system $\mathcal S$ with Hamiltonian $H_\mathcal S$ and initial state $\rho_\mathcal S$, coupled to a quantum memory $\mathcal A$ initially prepared in a pure ready state $\ket{0}$. The initial joint state is
\begin{equation}
\rho_\mathcal{SA}(0)=\rho_\mathcal{S}\otimes\ket0\!\bra0.
\label{eq:init_state}
\end{equation}
The demon protocol consists of two coherent stages. First, a premeasurement interaction correlates the system with an orthonormal memory basis $\{\ket a\}_{a=0}^d$ through a unitary operation $U_{\rm PM}$. Subsequently, a feedback operation conditioned on the memory state is implemented through
\begin{equation}
U_{\rm FB}=\sum_a U_a\otimes|a\rangle\!\langle a|,
\label{eq:U_FB}
\end{equation}
where each $U_a$ acts unitarily on the system. Throughout the protocol, the joint evolution of system and memory remains unitary; no measurement readout, postselection, classical feedback, or initial system--memory correlations are required.

The premeasurement generates correlations between system and memory that can subsequently be exploited through conditional control. The resulting joint evolution reads
\begin{equation}
\rho'_{\mathcal{SA}}=U_{\rm PM}\,\rho_{\mathcal{SA}}(0)\,U_{\rm PM}^{\dagger},
\end{equation}
followed by
\begin{equation}
\rho''_{\mathcal{SA}}=U_{\rm FB}\,
\rho'_{\mathcal{SA}}\,U_{\rm FB}^{\dagger}.
\end{equation}

The extracted work is quantified by the reduction of the system mean energy,
\begin{equation}
W=\Tr(H_\mathcal S\rho_\mathcal{S})-\Tr(H_\mathcal{S}\rho''_\mathcal{S}),
\label{eq:work_def}
\end{equation}
where $\rho''_\mathcal S=\Tr_\mathcal{A}(\rho''_{\mathcal{SA}})$. Throughout this work, the system Hilbert space has dimension $d$. For the protocol considered here, the memory is taken to have dimension $d+1$, consisting of one ready state $\ket0$ and $d$ conditional memory states $\{\ket a\}_{a=1}^{d}$ associated with the nontrivial branches of the protocol.

The coherent premeasurement induces an effective quantum instrument on the system when the memory degrees of freedom are ignored. The resulting reduced dynamics can be written as
\begin{equation}
\rho'_\mathcal{S}=\sum_{a=0}^d
M_a\rho_\mathcal{S} M_a^\dagger,
\label{eq:instrument}
\end{equation}
with Kraus operators
\begin{equation}
M_a=\bra a U_{\rm PM}\ket 0,
\end{equation}
satisfying $\sum_aM_a^\dagger M_a=\mathbbm{1}_\mathcal{S}$.

To investigate the weak-premeasurement regime, we consider
\begin{equation}
M_0=\sqrt{1-\varepsilon}\,\mathbbm 1_\mathcal S,\qquad M_a=\sqrt{\varepsilon}\,K_a \quad(a\ge1),
\label{eq:weak_Kraus}
\end{equation}
where $0\le\varepsilon\le1$ parametrizes the premeasurement strength, the weak-premeasurement regime corresponding to $\varepsilon<1$ (see Ref.~\cite{Dieguez2018} for a related monitoring framework), and $\sum_{a\ge1}K_a^\dagger K_a=\mathbbm{1}_\mathcal S$. Here and throughout, the index $a$ runs over the nontrivial conditional branches.

The corresponding conditional states and probabilities are
\begin{equation}
\rho_a=\frac{K_a\rho_\mathcal SK_a^\dagger}{p_a},
\qquad p_a=\Tr(K_a\rho_\mathcal SK_a^\dagger).
\label{eq:conditional_states}
\end{equation}
After the coherent feedback operation~\eqref{eq:U_FB}, the reduced system state becomes
\begin{equation}
\rho''_\mathcal{S} = (1-\varepsilon) U_0\rho_\mathcal{S}U_0^\dagger +\varepsilon
\sum_{a\ge1}p_a\,U_a\rho_aU_a^\dagger.
\label{eq:rhoS_after_FB}
\end{equation}
Substituting Eq.~\eqref{eq:rhoS_after_FB} into Eq.~\eqref{eq:work_def} yields
\begin{equation}
W_\varepsilon =(1-\varepsilon)W_0+\varepsilon W_{\rm dem},
\label{eq:W_convex}
\end{equation}
where
\begin{equation}
W_0 = \Tr(H_\mathcal S\rho_\mathcal S) - \Tr(H_\mathcal S U_0\rho_\mathcal S U_0^\dagger)
\end{equation}
is the work extracted through unconditional unitary driving, and
\begin{equation}
W_{\rm dem} = \Tr(H_\mathcal S\rho_\mathcal S)-\sum_{a\ge1} p_a\,
\Tr(H_\mathcal S U_a\rho_aU_a^\dagger)
\label{eq:W_dem_def}
\end{equation}
is the demon-assisted work contribution.

For the optimal choice of $U_0$, the quantity $W_0$ coincides with the standard ergotropy of the initial state~\cite{Allahverdyan2004}. Equation~\eqref{eq:W_convex} therefore establishes a continuous interpolation between standard ergotropy and fully resolved demon-assisted work extraction. In the weak-premeasurement regime, the resulting gain is naturally interpreted as a weak-ergotropic enhancement.

To identify the contribution enabled specifically by conditional control, we introduce the average conditional state
\begin{equation}
\bar{\rho}=\sum_{a\ge1}p_a\rho_a=
\sum_{a\ge1} K_a\rho_\mathcal S K_a^\dagger.
\label{eq:rho_bar}
\end{equation}
Using $\bar{\rho}$, Eq.~\eqref{eq:W_dem_def} can be decomposed as
\begin{equation}
W_{\rm dem}=W_{\rm PM}+W_{\rm FB},
\label{eq:Wdem_split}
\end{equation}
where
\begin{subequations}
\begin{align}
W_{\rm PM}&=\Tr(H_\mathcal{S}\rho_\mathcal{S})-
\Tr(H_\mathcal{S}\bar{\rho}),\\
W_{\rm FB}&=\Tr(H_\mathcal{S}\bar{\rho})-\sum_{a\ge1} p_a \Tr(H_\mathcal{S} U_a\rho_aU_a^\dagger).
\end{align}
\end{subequations}
The first contribution captures the energetic effect of the coherent premeasurement, whereas the second quantifies the gain achievable through feedback. However, part of $W_{\rm FB}$ is already accessible through optimal unconditional driving applied directly to the averaged state $\bar{\rho}$, without resolving the conditional branches. To isolate the advantage enabled specifically by branch-resolved control, we introduce the passive energy~\cite{PuszWoronowicz1978,Lenard1978,Allahverdyan2004}
\begin{equation}
\mathcal{E}(\sigma)=\min_U
\Tr(H_\mathcal S U\sigma U^\dagger),
\label{eq:passive_energy}
\end{equation}
which corresponds to the minimum energy attainable from
$\sigma$ under arbitrary unitary control. Consequently,
optimization over the feedback unitaries replaces the energy of each conditional branch by its passive energy. We therefore define
\begin{equation}
W_{\rm cc}=\mathcal{E}(\bar{\rho})-\sum_{a\ge1}
p_a\mathcal E(\rho_a),
\label{eq:Wcc}
\end{equation}
which quantifies the thermodynamic value of conditional control (hence the subscript ``cc''). More specifically, it measures the additional work enabled by resolving the conditional branches and applying branch-dependent control operations. By concavity of the passive energy, $W_{\rm cc}\ge0$, showing that conditional control can never reduce the optimal extractable work. 

The relevance of $W_{\rm cc}$ becomes apparent when comparing the optimal weak-ergotropic gain with the standard ergotropy of the initial state. Optimizing Eq.~\eqref{eq:W_convex} over the available unitary controls $\{U_a\}$ and $U_0$ yields the maximal work extraction, denoted throughout by the superscript $\star$. Adding and subtracting $\mathcal E(\bar\rho)$ then gives
\begin{align}
\delta W_\varepsilon^\star &=W_\varepsilon^\star-W_0^\star\nonumber\\ 
&=\varepsilon\Big[
\mathcal{E}(\rho_\mathcal{S})-\mathcal{E}(\bar\rho)+ W_{\rm cc}\Big].
\label{eq:dW}
\end{align}
The first term reflects the energetic effect of the average-state transformation induced by the coherent premeasurement, whereas $W_{\rm cc}$ quantifies the additional advantage enabled by conditional control. While $\delta W_\varepsilon^\star$ captures the overall demon-induced enhancement, $W_{\rm cc}$ isolates the contribution that is exclusively associated with branch-resolved control.

The central question addressed in this Letter can now be formulated precisely: what determines the thermodynamic value of conditional control, as quantified by $W_{\rm cc}$? The remainder of this work is devoted to answering this question.

\textit{Information versus thermodynamic value.}---Since $W_{\rm cc}$ is enabled by resolving conditional branches, it is natural to compare it with informational descriptors of the ensemble $\{p_a,\rho_a\}_{a\ge1}$. We consider the Holevo information
\begin{equation}
\chi=S(\bar{\rho})-\sum_{a\ge1}p_a S(\rho_a),
\label{eq:Holevo}
\end{equation}
and the accessible trace-distance measure
\begin{equation}
T_{\rm acc}=\sum_{a\ge1}p_a\left\|\rho_a-\bar{\rho}\right\|_1.
\label{eq:Tacc}
\end{equation}
The former quantifies the information about the branch label that can in principle be encoded in the conditional ensemble, whereas the latter measures how far the branches are, on average, from the state obtained when the branch information is ignored.

These quantities constrain the conditional-control advantage. Indeed, using the variational characterization of passive energy (see Supplemental Material), one obtains
\begin{equation}
W_{\rm cc}\le\frac{\Delta H_\mathcal S}{2}\,T_{\rm acc},
\label{eq:Tacc_bound}
\end{equation}
where $\Delta H_\mathcal S=\epsilon_{\max}-\epsilon_{\min}$ is the spectral range of the system Hamiltonian. Combining this bound with Pinsker's inequality gives
\begin{equation}
W_{\rm cc}\le\frac{\Delta H_\mathcal S}{2}\sqrt{2\chi}.
\label{eq:Holevo_bound}
\end{equation}
Equations~\eqref{eq:Tacc_bound} and~\eqref{eq:Holevo_bound}
show that information constrains the thermodynamic value of
conditional control. In particular, if the conditional branches carry no distinguishable information, then $W_{\rm cc}$ must vanish.

However, these bounds do not imply that information determines the thermodynamic value of conditional control. They only establish that distinguishability is necessary for a branch-resolved advantage to exist. By contrast, the total weak-ergotropic enhancement $\delta W_\varepsilon^\star$ introduced in Eq.~\eqref{eq:dW} may remain strictly positive even when
\begin{equation}
T_{\rm acc}=0,\qquad\chi=0,\qquad W_{\rm cc}=0,
\end{equation}
showing that information is not the sole ingredient underlying the overall demon-induced gain \footnote{A simple example is provided by conditional ensembles satisfying $\rho_a=\bar\rho$ for every $a$. In this case $T_{\rm acc}=\chi=W_{\rm cc}=0$, while $\mathcal{E}(\rho_\mathcal{S})-\mathcal{E}(\bar\rho)$ may remain positive whenever $\bar\rho\neq\rho_\mathcal{S}$, yielding a strictly positive weak-ergotropic enhancement $\delta W_\varepsilon^\star$.}. The more fundamental question concerns $W_{\rm cc}$ itself. Do $\chi$ and $T_{\rm acc}$ determine the conditional-control advantage once the average state and the Hamiltonian are fixed? The answer is negative. As shown explicitly in the Supplemental Material, one can construct two conditional ensembles associated with the same system Hamiltonian such that
\begin{equation}
\bar\rho_1=\bar\rho_2,\qquad\chi_1=\chi_2,\qquad T_{{\rm acc},1}=T_{{\rm acc},2},
\label{eq:same_info}
\end{equation}
while nevertheless
\begin{equation}
W_{{\rm cc},1}\neq W_{{\rm cc},2}.
\label{eq:different_Wcc}
\end{equation}
Hence, even after fixing the average state, the Holevo information, and the distinguishability of the conditional ensemble, the thermodynamic value of conditional control remains undetermined.

This failure of scalar information measures reveals the
missing ingredient. Information quantifies how well conditional branches can be resolved, but the energetic value of resolving them depends on how their spectra are rearranged relative to the passive ordering imposed by the Hamiltonian. We now show that this missing structure admits a natural geometric representation in terms of passive spectral rearrangements and Hamiltonian gap structures.

\textit{Passive spectral rearrangement.}---The previous discussion shows that information constrains the thermodynamic value of conditional control but does not determine it. We now identify the quantity that does. The key observation is that $W_{\rm cc}$ is defined in terms of passive energies. Let
\begin{equation}
H_\mathcal{S} =\sum_{i=1}^{d}\epsilon_i \Pi_i,\qquad
\epsilon_1\le\epsilon_2\le\cdots\le\epsilon_d,
\end{equation}
and denote by $\lambda_i^\downarrow(\rho)$
the eigenvalues of $\rho$ arranged in nonincreasing order. The passive energy can then be written as
\begin{equation}
\mathcal{E}(\rho)=\sum_{i=1}^{d}\epsilon_i
\lambda_i^\downarrow(\rho).
\label{eq:passive_spectral}
\end{equation}
Substituting Eq.~\eqref{eq:passive_spectral} into the definition of
$W_{\rm cc}$ gives
\begin{equation}
W_{\rm cc}=\sum_{i=1}^{d}\epsilon_i
\left[\lambda_i^\downarrow(\bar\rho)-\sum_{a\ge1} p_a
\lambda_i^\downarrow(\rho_a)\right].
\label{eq:Wcc_spectral}
\end{equation}
At first sight, Eq.~\eqref{eq:Wcc_spectral} appears to depend on the individual ordered eigenvalues. However, a more revealing structure emerges when these quantities are reorganized into cumulative spectral weights. To this end, we introduce 
\begin{equation}
\Lambda_k(\rho)=\sum_{i=1}^{k}\lambda_i^\downarrow(\rho),
\qquad k=1,\dots,d-1,
\label{eq:KyFan}
\end{equation}
which coincide with the Ky Fan spectral sums~\cite{KyFan1949,Bhatia1997}. Using Eq.~\eqref{eq:KyFan}, we define
\begin{equation}
R_k=\sum_{a\ge1} p_a \Lambda_k(\rho_a)-\Lambda_k(\bar\rho).
\label{eq:Rk}
\end{equation}
Since the Ky Fan sums are convex functions of the state,
\begin{equation}
\Lambda_k(\bar\rho)=\Lambda_k\!\left(\sum_a p_a\rho_a\right)\le \sum_a p_a \Lambda_k(\rho_a),
\end{equation}
and therefore
\begin{equation}
R_k\ge0.
\label{eq:Rk_positive}
\end{equation}
Each component $R_k$ quantifies how much the conditioning process increases the average weight contained in the $k$ largest eigenvalues relative to the average state. Collectively, the vector
\begin{equation}
\mathbf{R}=(R_1,\ldots,R_{d-1})
\end{equation}
provides a complete characterization of the passive spectral rearrangement induced by conditioning. A straightforward summation-by-parts manipulation of Eq.~\eqref{eq:Wcc_spectral} yields our main result:
\begin{equation}
W_{\rm cc}=\sum_{k=1}^{d-1}(\epsilon_{k+1}-\epsilon_k)
R_k.
\label{eq:main_theorem}
\end{equation}
Defining the energy-gap vector
\begin{equation}
\boldsymbol{\Delta\epsilon}=(\epsilon_2-\epsilon_1,\dots,
\epsilon_d-\epsilon_{d-1}),
\end{equation}
Eq.~\eqref{eq:main_theorem} can be written compactly as
\begin{equation}
W_{\rm cc}=\boldsymbol{\Delta\epsilon}\cdot \mathbf{R}.
\label{eq:vector_form}
\end{equation}
Equivalently, for $\|\mathbf{R}\|\neq0$ and $\|\boldsymbol{\Delta\epsilon}\|\neq0$,
\begin{equation}
W_{\rm cc}=\|\boldsymbol{\Delta\epsilon}\|\,\|\mathbf{R}\|\cos\theta,
\label{eq:angle_form}
\end{equation}
where $\theta$ denotes the angle between the energy-gap vector and the passive spectral rearrangement vector. 

Equation~\eqref{eq:vector_form} reveals the thermodynamic geometry of conditional control. The Hamiltonian enters exclusively through its gap structure $\boldsymbol{\Delta\epsilon}$, whereas the conditional ensemble contributes exclusively through the passive spectral rearrangement vector $\mathbf R$. The conditional-control advantage is therefore not a property of either object individually, but rather of their geometric pairing.

This result also explains the informational ambiguity discussed in the previous section. The counterexample is not accidental: scalar informational quantities such as the Holevo information and the accessible distinguishability do not uniquely specify the passive spectral rearrangement induced by conditioning. Consequently, they cannot uniquely determine the thermodynamic value of conditional control.

\textit{Physical interpretation.}---The thermodynamic geometry identified by Eq.~\eqref{eq:vector_form} admits a direct physical interpretation. The passive spectral rearrangement vector $\mathbf{R}$ characterizes the effect of conditioning on the ensemble, whereas the energy-gap vector $\boldsymbol{\Delta\epsilon}$ characterizes the energetic structure of the Hamiltonian.

This geometric character becomes particularly transparent through Eq.~\eqref{eq:angle_form}. The quantity $\|\mathbf{R}\|$ measures the overall amount of passive spectral rearrangement generated by conditioning, whereas the factor $\cos\theta$ quantifies the compatibility between this rearrangement and the Hamiltonian gap structure. Large passive rearrangements need not produce a large thermodynamic value if they are poorly aligned with the energetic structure of the system. Conversely, a modest rearrangement can become thermodynamically valuable when it is strongly aligned with the Hamiltonian gaps.

The components of $\mathbf{R}$ provide a finer description of this geometry. Each component $R_k$ quantifies the excess average weight accumulated in the $k$ largest eigenvalues of the conditional states relative to the average state. The vector $\mathbf{R}$ therefore quantifies a complete hierarchy of cumulative passive spectral rearrangements induced by conditioning. Equation~\eqref{eq:main_theorem} shows that the Hamiltonian assigns a distinct energetic value to each of these rearrangements through the corresponding energy gap $(\epsilon_{k+1}-\epsilon_k)$.

An especially illuminating situation arises for a two-level system. In this case there is only one independent cumulative spectral weight,
\begin{equation}
R_1=\sum_a p_a\lambda_1^\downarrow(\rho_a)-\lambda_1^\downarrow(\bar\rho),
\end{equation}
and Eq.~\eqref{eq:main_theorem} reduces to
\begin{equation}
W_{\rm cc}=(\epsilon_2-\epsilon_1)R_1.
\label{eq:qubit_case}
\end{equation}
The thermodynamic geometry therefore becomes one-dimensional: there is a single passive spectral rearrangement and a single energy gap, so no nontrivial alignment structure can arise. Higher-dimensional systems exhibit a richer geometry, with multiple independent rearrangements contributing to the conditional-control advantage through their interplay with the Hamiltonian gap structure.

Equation~\eqref{eq:vector_form} thus identifies the geometric ingredient that information-theoretic measures fail to capture. While distinguishability quantifies how well conditional branches can be resolved, the thermodynamic value of resolving them is determined by the geometric relation between passive spectral rearrangements and Hamiltonian gap structures. This distinction becomes explicit in the example discussed below.

\textit{Informational equivalence and thermodynamic value.}---The counterexample discussed in the Supplemental Material acquires a transparent interpretation in light of Eq.~\eqref{eq:vector_form}. Two conditional ensembles may share the same average state and identical informational descriptors, such as $\chi$ and $T_{\rm acc}$, while possessing different passive spectral rearrangements. In that case,
$\mathbf{R}^{(1)}\neq \mathbf{R}^{(2)}$, and therefore
$W_{{\rm cc},1}=\boldsymbol{\Delta\epsilon}\cdot \mathbf{R}^{(1)}\neq\boldsymbol{\Delta\epsilon}\cdot \mathbf{R}^{(2)}=W_{{\rm cc},2}$.

This result highlights the distinction between informational and thermodynamic descriptions of conditional control. Informational quantities characterize how well conditional branches can be resolved, whereas the vector $\mathbf{R}$ characterizes their thermodynamic geometry. Two ensembles may therefore be informationally equivalent while occupying distinct positions in the geometric framework defined by Eq.~\eqref{eq:vector_form}.

The ambiguity left unresolved by informational quantities is thus completely removed by the passive spectral rearrangement vector. Informational equivalence does not imply thermodynamic equivalence.

\textit{Conclusion.}---We investigated the thermodynamic value of conditional control within a fully coherent demon protocol and answered the central question posed in the introduction: what determines its thermodynamic value? Although informational descriptors such as the Holevo quantity and accessible distinguishability constrain the conditional-control advantage, they do not uniquely determine it. Informationally equivalent ensembles may possess different thermodynamic values.

We showed that the missing descriptor is the passive spectral rearrangement induced by conditioning. The conditional-control advantage is completely characterized by the relation
$W_{\rm cc}=\boldsymbol{\Delta\epsilon}\cdot \mathbf{R}$, where the passive spectral rearrangement vector $\mathbf{R}$ encodes the effect of conditioning on the ensemble and the energy-gap vector $\boldsymbol{\Delta\epsilon}$ encodes the energetic structure of the Hamiltonian. This result reveals the thermodynamic geometry of conditional control and establishes a clear distinction between information and thermodynamic value. Information enables conditional control, but its thermodynamic value is determined by how passive spectral rearrangements are valued by the Hamiltonian gap structure. In this sense, the passive spectral rearrangement vector $\mathbf{R}$ provides the minimal operational descriptor needed to determine the thermodynamic value of conditional control for a fixed Hamiltonian.

The thermodynamic closure of the complete demon cycle, including memory-resetting costs consistent with Landauer's principle, is discussed in the Supplemental Material. Our framework also provides a natural setting for what may be viewed as a weak-ergotropic regime, in which demon-assisted work extraction emerges continuously from a coherent premeasurement process without measurement readout or classical feedback. In this perspective, the resulting weak-ergotropic enhancement appears not as an independent resource, but as a thermodynamic manifestation of the passive spectral rearrangements generated during conditional control.

Beyond its conceptual implications, the present framework provides a new way of characterizing conditional-control resources that goes beyond conventional information-theoretic measures. More generally, the emergence of a thermodynamic geometry based on passive spectral rearrangements and Hamiltonian gap structures opens new perspectives for the study of information-assisted work extraction, coherent feedback protocols, and majorization-based approaches to quantum thermodynamics. An interesting direction for future work is the characterization of the space of physically admissible passive spectral rearrangements and its relation to thermodynamic resource theories.

\textit{Acknowledgments}. We thank Pedro Ruas Dieguez and Tiago Debarba for carefully reading an earlier version of this manuscript and for valuable discussions. J.G.G.O.Jr. acknowledges support from CAPES (Coordenação de Aperfeiçoamento de Pessoal de Nível Superior – Brazil), Grant No.~88887.909640/2023-00. R.M.A. and A.C.S.C. acknowledges support from the Conselho Nacional de Desenvolvimento Científico e Tecnológico (CNPq), Grants No.~305957/2023-6 and 308730/2023-2, respectively.

\textit{Data availability}---This work is theoretical in nature and does not rely on datasets or software. All results are derived analytically and are fully contained within the manuscript and Supplemental Material.

\appendix
\section{\bf SUPPLEMENTAL MATERIAL}

\section{I. Derivation of Eq.~\eqref{eq:main_theorem} of the main text}
\label{SM:theorem}

Here we derive Eq.~\eqref{eq:main_theorem} from Eq.~\eqref{eq:Wcc_spectral} of the main text. Let
\begin{equation}
D_i=\lambda_i^\downarrow(\bar\rho)-\sum_a p_a\lambda_i^\downarrow(\rho_a),
\end{equation}
so that Eq.~\eqref{eq:Wcc_spectral} can be written as
\begin{equation}
W_{\rm cc}=\sum_{i=1}^{d}\epsilon_i D_i .
\label{SM:Wcc_D}
\end{equation}
Define the cumulative quantities
\begin{equation}
S_k=\sum_{i=1}^{k}D_i,\qquad k=1,\ldots,d-1.
\label{SM:Sk}
\end{equation}
Since both $\bar\rho$ and the conditional states are normalized,
\begin{equation}
\sum_{i=1}^{d}D_i=0.
\end{equation}
Applying the discrete summation-by-parts identity to Eq.~\eqref{SM:Wcc_D} yields
\begin{equation}
W_{\rm cc}=\sum_{k=1}^{d-1}(\epsilon_k-\epsilon_{k+1})S_k .
\label{SM:Wcc_Sk}
\end{equation}
Using Eq.~\eqref{eq:Rk} of the main text,
\begin{equation}
S_k=\Lambda_k(\bar\rho)-\sum_a p_a \Lambda_k(\rho_a)=-R_k,
\end{equation}
and therefore
\begin{equation}
W_{\rm cc}=\sum_{k=1}^{d-1}(\epsilon_{k+1}-\epsilon_k)R_k,
\end{equation}
which is Eq.~\eqref{eq:main_theorem} of the main text.

\section{II. Information-theoretic bounds on conditional control}
\label{SM:bounds}

In this section we derive the bounds stated in Eqs.~\eqref{eq:Tacc_bound} and \eqref{eq:Holevo_bound} of the main text. Starting from Eq.~\eqref{eq:Wcc}, $W_{\rm cc}=\mathcal E(\bar\rho)-\sum_a p_a \mathcal E(\rho_a)$, we write
\begin{align}
W_{\rm cc}&=\sum_a p_a\Big[\mathcal{E}(\bar\rho)-\mathcal{E}(\rho_a)\Big].
\label{SM:bound_start}
\end{align}
The passive energy admits the variational representation 
\begin{equation}
\mathcal{E}(\rho)=\min_U\Tr(H_{\mathcal S}U\rho U^\dagger).
\label{SM:variational_passive}
\end{equation}
For arbitrary states $\rho$ and $\sigma$, let $U_\sigma$ be a unitary achieving the minimum in Eq.~\eqref{SM:variational_passive} for $\sigma$. Then
\begin{align}
\mathcal E(\rho)-\mathcal E(\sigma)&=\mathcal E(\rho)-\Tr(H_{\mathcal S}U_\sigma\sigma U_\sigma^\dagger)
\nonumber\\
&\le\Tr\!\left[H_{\mathcal S}U_\sigma(\rho-\sigma)U_\sigma^\dagger\right].
\end{align}
Using the Hölder inequality
\begin{equation}
|\Tr(XY)|
\le
\|X\|_\infty\|Y\|_1,
\end{equation}
where $\|\cdot\|_1$ and $\|\cdot\|_\infty$ denote the trace and operator norms, respectively, one obtains
\begin{equation}
\mathcal{E}(\rho)-\mathcal{E}(\sigma)\le
\|H_{\mathcal{S}}\|_\infty\,\|\rho-\sigma\|_1.
\end{equation}
A tighter estimate follows by shifting the Hamiltonian by a multiple of the identity. Since passive energies are invariant under the transformation $H_\mathcal{S}\mapsto H_\mathcal{S}-c\,\mathbbm{1}$, we choose $c=(\epsilon_{\max}+\epsilon_{\min})/2$. The resulting operator satisfies
\begin{equation}
\left\|H_{\mathcal S}-c\,\mathbbm 1\right\|_\infty=
\frac{\Delta H_{\mathcal S}}{2},
\end{equation}
where $\Delta H_{\mathcal S}=\epsilon_{\max}-\epsilon_{\min}$. Therefore,
\begin{equation}
\mathcal E(\rho)-\mathcal E(\sigma)\le\frac{\Delta H_{\mathcal S}}{2}\,\|\rho-\sigma\|_1.
\label{SM:Lipschitz}
\end{equation}
Applying Eq.~\eqref{SM:Lipschitz} to Eq.~\eqref{SM:bound_start} gives
\begin{align}
W_{\rm cc}&\le\frac{\Delta H_{\mathcal S}}{2}\sum_a p_a\|\rho_a-\bar\rho\|_1=\frac{\Delta H_{\mathcal S}}{2}
\,T_{\rm acc},
\end{align}
which proves Eq.~\eqref{eq:Tacc_bound}.

To obtain the Holevo bound, we use the quantum Pinsker inequality,
\begin{equation}
\|\rho-\sigma\|_1
\le
\sqrt{2\,D(\rho\|\sigma)},
\end{equation}
where $D(\rho\|\sigma)=\Tr[\rho(\ln\rho-\ln\sigma)]$ is the quantum relative entropy, together with the concavity of the square root:
\begin{align}
T_{\rm acc}&=\sum_a p_a\|\rho_a-\bar\rho\|_1\nonumber\\
&\le\sum_a p_a\sqrt{2\,D(\rho_a\|\bar\rho)}\nonumber\\
&\le\sqrt{2\sum_a p_aD(\rho_a\|\bar\rho)}.
\end{align}
With the identity $\chi=\sum_a p_a D(\rho_a\|\bar\rho)$ one finds $T_{\rm acc}\le\sqrt{2\chi}$. Combining this relation with Eq.~\eqref{eq:Tacc_bound} yields
\begin{equation}
W_{\rm cc}
\le
\frac{\Delta H_{\mathcal S}}{2}
\sqrt{2\chi},
\end{equation}
which proves Eq.~\eqref{eq:Holevo_bound}.

\section{III. Informationally equivalent ensembles with different thermodynamic values}
\label{SM:example}

We present an explicit qutrit example showing that $\bar\rho$, $\chi$, and $T_{\rm acc}$ do not determine $W_{\rm cc}$. Since the thermodynamic geometry developed in the main text depends only on the conditional ensemble, it is sufficient to construct the example directly in terms of the states $\rho_a$ and probabilities $p_a$. Consider the average state
\begin{equation}
\bar\rho=\mathrm{diag}\left(\frac12,\frac{3}{10},\frac15\right),
\end{equation}
and two equiprobable conditional branches, $p_1=p_2=1/2$, defined by
\begin{subequations}
\begin{align}
\rho_1(\nu)&=\mathrm{diag}\left(\frac14,\frac{3}{10}+\nu,\frac{9}{20}-\nu\right),\\
\rho_2(\nu)&=\mathrm{diag}\left(\frac34,\frac{3}{10}-\nu,\nu-\frac{1}{20}\right).
\end{align}
\end{subequations}
For $\nu\in[3/40,7/40]$, these are valid density matrices and satisfy
\begin{equation}
\frac12\rho_1(\nu)+\frac12\rho_2(\nu)=\bar\rho.
\end{equation}
Moreover,
\begin{equation}
T_{\rm acc}(\nu)=\frac12
\end{equation}
throughout the interval.

The Holevo quantity can be evaluated analytically from the spectra of the conditional states,
\begin{equation}
\chi(\nu)=S(\bar\rho)-\frac12S\big(\rho_1(\nu)\big)-\frac12S\big(\rho_2(\nu)\big).
\end{equation}
It possesses a unique minimum at
\begin{equation}
\nu_0=\frac{3}{20},
\end{equation}
around which
\begin{equation}
\chi(\nu)=\log\!\left(\frac{3^{3/4}}{2}\right)
+\frac{50}{9}\left(\nu-\nu_0\right)^2
+O\!\left[\left(\nu-\nu_0\right)^3\right].
\end{equation}
Since $\chi(\nu)$ is continuous and has a strict minimum at $\nu_0$, every value of $\chi$ sufficiently close to the minimum is attained at two distinct values of $\nu$ lying on opposite sides of $\nu_0$. Writing these values as
\begin{equation}
\nu_\pm=\nu_0\pm\delta,
\end{equation}
one has $\chi(\nu_+)=\chi(\nu_-)$. The local expansion above illustrates the origin of this degeneracy.

On the other hand, throughout this interval the passive spectral rearrangement vector is
\begin{equation}
\mathbf{R}(\nu)=\left(\frac{1}{40}+\frac{\nu}{2},\frac{1}{10}-\frac{\nu}{2}\right).
\end{equation}
To make the thermodynamic inequivalence explicit, consider a qutrit Hamiltonian with gap vector
\begin{equation}
\boldsymbol{\Delta\epsilon}=\epsilon_0(1,2),
\end{equation}
where $\epsilon_0$ sets the energy scale. Using Eq.~\eqref{eq:vector_form}, one finds
\begin{equation}
W_{\rm cc}(\nu)=\boldsymbol{\Delta\epsilon}\cdot \mathbf{R}(\nu)=\epsilon_0\left(\frac{9}{40}-\frac{\nu}{2}\right).
\end{equation}
Therefore,
\begin{equation}
\chi(\nu_+)-\chi(\nu_-)=O(\delta^3),
\end{equation}
whereas
\begin{equation}
W_{\rm cc}(\nu_+)-W_{\rm cc}(\nu_-)=-\epsilon_0\,\delta.
\end{equation}
This demonstrates that conditional ensembles may become arbitrarily close from the informational perspective while remaining thermodynamically distinguishable. The passive spectral rearrangement therefore constitutes a thermodynamic descriptor of conditional control that is not captured by standard informational quantities such as $\chi$ and $T_{\rm acc}$.

\section{IV. Thermodynamic closure and Landauer cost}
\label{SM:Landauer}

The quantity $\delta W_\varepsilon^\star$ introduced in Eq.~\eqref{eq:dW} characterizes the work enhancement enabled by the coherent demon protocol. However, it does not represent the net work obtainable from a complete thermodynamic cycle. To restore the demon to its initial ready state, the memory must eventually be reset.

Let $\rho_{\mathcal A}''=\Tr_{\mathcal S}(\rho_{\mathcal{SA}}'')$ denote the final state of the memory after completion of the coherent protocol. If the memory is reset locally through contact with a thermal reservoir at temperature $T$, Landauer's principle implies the work cost
\begin{equation}
W_{\rm reset}\ge k_B T\,S(\rho_{\mathcal A}''),
\end{equation}
where $S(\rho_{\mathcal A}'')$ is the von Neumann entropy of the memory state~\cite{Landauer,ReebWolf2014}. Considering only the demon-induced enhancement, one obtains
\begin{equation}
W_{\rm net}\le\delta W_\varepsilon^\star-k_B T\,S(\rho_{\mathcal A}'').
\end{equation}

A more refined description becomes possible when the residual correlations between system and memory are also taken into account during the resetting process. In this case, the resetting cost can, in principle, be characterized in terms of the conditional entropy
\begin{equation}
S(\mathcal A|\mathcal S)=S(\rho_{\mathcal{SA}}'')-S(\rho_{\mathcal S}''),
\end{equation}
and may be reduced accordingly~\cite{delRio2011}. The difference between the local and conditional resetting costs is determined by the mutual information
\begin{equation}
I(\mathcal{S\!:\! A})=S(\rho_{\mathcal S}'')+S(\rho_{\mathcal A}'')-S(\rho_{\mathcal{SA}}''),
\end{equation}
through the identity
\begin{equation}
S(\rho_{\mathcal A}'')=S(\mathcal A|\mathcal S)+I(\mathcal{S\!:\!A}).
\end{equation}
Accordingly, if the resetting protocol is allowed to exploit the residual system--memory correlations, the informational contribution to the thermodynamic cost may be reduced from
$k_BT\,S(\rho_{\mathcal A}'')$ to $k_BT\,S(\mathcal A|\mathcal S)$, thereby increasing the net work obtainable from the complete cycle by an amount bounded by $k_BT\,I(\mathcal S\!:\!\mathcal A)$.

These observations highlight an important conceptual distinction. The memory-resetting cost is governed by informational quantities associated with the final system--memory state, whereas the useful contribution associated with conditional control is governed by the thermodynamic geometry of the conditional ensemble through $W_{\rm cc}=\boldsymbol{\Delta\epsilon}\cdot \mathbf{R}$. Information therefore governs the thermodynamic closure of the cycle, while passive spectral rearrangement governs the thermodynamic value of conditional control.

More generally, the irreversibility of the complete coherent demon cycle is linked to the residual system--memory correlations generated during the protocol. Thus, although information and thermodynamic geometry both play essential roles, they characterize distinct aspects of the demon's operation.

\bibliographystyle{apsrev4-2}
\bibliography{refs}

\end{document}